\documentclass{aastex} 
\usepackage{spr-astr-addons} 

\usepackage{graphicx, color}
\usepackage{amsmath}
\begin{document}

%
\title{Modelling of the radio spectrum evolution in the binary pulsar B1259-63}

\shorttitle{Modelling of the PSR B1259-63 radio spectrum evolution}
\shortauthors{Dembska et al.}

\author{M. Dembska\altaffilmark{1}}
\email{marta.dembska@dlr.de} 
\and 
\author{J. Kijak\altaffilmark{2}}
\and
\author{O. Koralewska\altaffilmark{2}}
\and
\author{W. Lewandowski\altaffilmark{2}}
\and
\author{G. Melikidze\altaffilmark{2,3}}
\and
\author{K. Ro\.{z}ko\altaffilmark{2}}

\altaffiltext{1}{German Aerospace Center, Institute of Space Systems, Robert Hooke Str. 7, 28359 Bremen, Germany}
\altaffiltext{2}{Institute of Astronomy, University of Zielona G\'ora, Lubuska 2, 65-265 Zielona G\'ora, Poland}
\altaffiltext{3}{Abastumani Astrophysical Observatory, Ilia State University, 3-5 Cholokashvili Ave.,Tbilisi, 0160, Georgia}

\begin{abstract}
In this paper we give the first attempt to model the evolution of the spectrum of PSR B1259$-$63 radio emission while the pulsar orbits the companion Be star. As suggested by Kijak et al. (2011, MNRAS, 418, L114) this binary system can be useful in understanding the origin of the gigahertz-peaked spectrum of pulsars. The model explains, at least qualitatively, the observed alterations of the spectral shape depending on the orbital phases of this pulsar. Thus, our results support the hypothesis that the external factors have a significant impact on the observed radio emission of a pulsar. The model can also contribute to our understanding of the origin of some non-typical spectral shapes (e.g. flat or broken spectra).
\end{abstract}

\section{Introduction}

The gigahertz-peaked spectra (GPS) pulsars were introduced by \citet{kijak2011b} providing a definite evidence for a new type of radio spectra of pulsars whose flux density reaches the maximum value at frequencies above 1 GHz. At higher frequencies the spectra of these pulsars look like a typical pulsar spectrum, while at frequencies below 1 GHz the observed flux density decreases showing a positive spectral index. The frequency at which such a spectrum reaches the highest value (i.e. the flux density reaches the maximum) was called the peak frequency $\nu_p$. \citet{kijak2011b} also indicated that the GPS pulsars are relatively young objects and they usually adjoin some active environments like HII regions or compact pulsar wind nebulae (PWNe).

On the other hand the pulsar spectral shape provides important information for understanding of the pulsar radio emission mechanism. The radio spectrum of typical pulsars can be described using either a simple power law with a negative spectral index of $-$1.8 or two power laws with spectral indices of $-0.9$ and $- 2.2$ and a break frequency on average of 1.5 GHz~\citep{maron2000}. Some pulsar show a low frequency turnover in their spectra\citep{sieber73, malofeev94}. \citet{lorimer95} also reported several pulsars having positive spectral index or flat spectrum in the frequency range between 400 MHz and 1600 MHz. \citet{kijak2007} presented the first direct evidence of a high-frequency turnover in radio pulsars spectra. \citet{kijak2011b} noticed that all the five GPS pulsars have a high value of dispersion measure ($DM$ around 150~$\mathrm{pc\,cm^{-3}}$ and above). However,  \citet{dembska2014} recently identified another two sources that exhibit a GPS spectrum, including a first low-DM GPS pulsar, PSR J1740+1000 which was reported to have a PWN in its surroundings. The GPS phenomenon was also revealed in two radio magnetars that assosiated with supernova remnants \citep{kijak2013} and another radio pulsar, PSR J2007+2722~\citep{einstein}.

\begin{figure*}[!t]
   \includegraphics[width=0.99\textwidth]{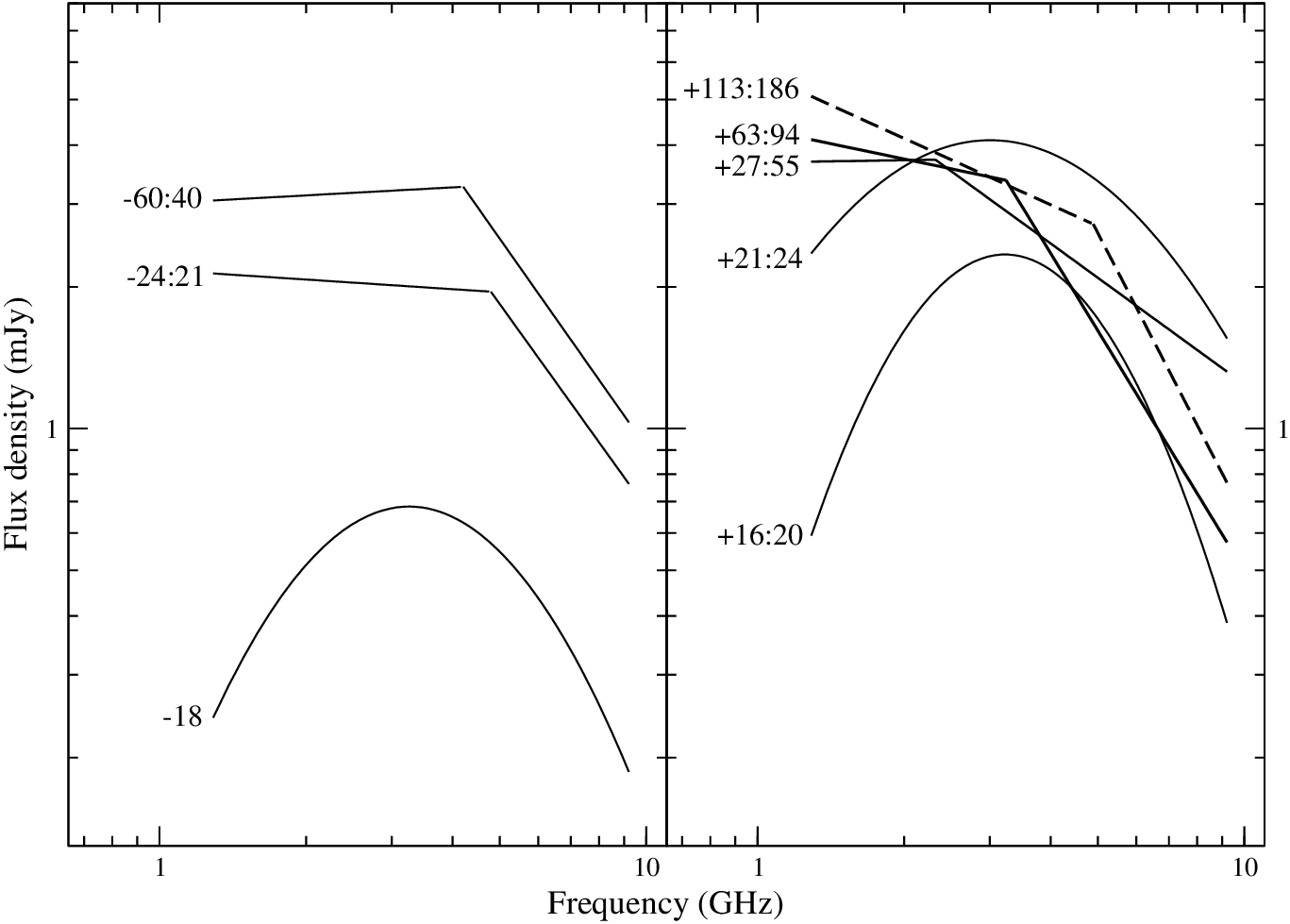}
   \caption{The fits to the PSR B1259$-$63 (J1302-6350) spectra for the each of the considered orbital phase intervals, prior to (left panel) and after (right panel) the periastron passage. See also, Section Discussion on the possible effects of scintillation}
     \label{intervals}
\end{figure*}

PSR B1259$-$63 (J1302$-$6350) was reported by \citet{lorimer95} as a pulsar with positive spectral index. This fact was the primary reason (along with radio spectrum evolution, age and value of dispersion measure) why this pulsar was considered as a good candidate for GPS pulsar. PSR B1259$-$63 is a member of the unique binary system along with a massive (10M$_{\odot}$) main-sequence Be star LS2883 with a radius of 6R$_{\odot}$ (here $M_{\odot}$ and $R_{\odot}$ are the solar mass and radius respectively). \citet{neug2011} estimates the mass of LS 2883 to be about 30M$_{\odot}$, however, their result is subject to large uncertainties. PSR B1259$-$63 is a middle-aged pulsar (330 kyr) with a short period of 48 ms. Its average DM is about 147 pc cm$^{-3}$. The orbital period of this eccentric ($e=0.87$) binary system is 3.4 yr and a projected semi-major axis $a\sin{i}$ is about 1300 light seconds (2.6 AU). \citet{johnston94} and \citet{melatos1995} suggested the existence of a disk around LS 2883. Both the star and its disk possibly produce strong magnetic field. It was shown that the disk is highly tilted with respect to the orbital plane by an angle 25$^\mathrm{o}$ and PSR B1259$-$63 is eclipsed for about 35 days as it goes behind the disk~\citep{johnston2005}. The PSR B1259$-$63/LS 2883 emits unpulsed non-thermal emission over a wide range of frequencies ranging from radio to  $\mathrm{\gamma}$-rays, and the fluxes vary with orbital phase.

\citet[Paper I, hereafter]{kijak2011a} studied the radio spectrum of PSR B1259$-$63 using the available measurements of the pulsed flux obtained during three periastron passages (1997, 2000 and 2004). They also presented spectral evolution of the pulsar radio emission due to its orbital motion and dependence of the peak frequency on the orbital phase. At the same time, they suggested that PSR B1259$-$63 can be treated as a key object for understanding the physical mechanisms which potentially are responsible for GPS phenomenon. \citet[Paper I]{kijak2011a} proposed two effects possibly responsible for the observed variation in spectra: the free-free absorption in the stellar wind and the cyclotron resonance in the magnetic field. This field is associated with the disk and is infused by the relativistic particles of the pulsar wind. The apparent resemblance between the PSR B1259$-$63 spectrum and the GPS suggests that the same mechanisms should be responsible for both cases (Paper I). Thus, there appears a conclusion that the GPS feature should be caused by some external factors rather than by the emission mechanism.

\begin{table}[!ht]
  \caption{The chosen fits to PSR B1259$-$63 spectra for each observing day, ``$-$" and ``+" denote days before and after periastron passage respectively, year is an epoch of observation. For more details on best fit marks see table comments.}
  \begin{tabular}{@{}clcl@{}}
\hline
   Day     &    Year     &  Best fit  & Comments \\
\hline
 -60 & 1997  & t & \\
 -48 & 1997  & u & no trend\\
 -46 & 2000  & b & \\
 -40 & 1997  & t  & \\
 -24 & 1997  & t  & \\
 -21 & 2004  & t  & \\
 -18 & 2000  & t & \\
+16 & 1997  & t  & \\
+20 & 1997  & t & \\
+21 & 2004  & b/t& small positive spectral index\\
        &           &                           & below 2 GHz\\
+24 & 2000  & t & \\
+27 & 1997  & t & \\
+28 & 2004  & p & \\
+29 & 2000  & p & \\
+30 & 1997  & u & no trend\\
+33 & 2004  & p/b& small break \\
+34 & 2000  & b/t & small positive spectral index\\
        &           &                           & below 4.2 GHz\\
+36 & 1997  & t & \\
+38 & 2004  & b/u & difficult to classify\\
+40 & 2000  & u & no trend\\
+42 & 1997  & f& \\
+47 & 1997  & b/t& small positive spectral index\\
        &           &                           & below 2.5 GHz\\
+55 & 2000  & u &  no trend\\
+63 & 2000  & b & \\
+64 & 2004  & t & \\
+70 & 2000  & p/b & small break\\
+83 & 2000  & b & \\
+94 & 2004  & b & \\
+113 & 2000  & u &  significant error bars\\
+150 & 2004  & b & \\
+186 & 2004  & p & \\
\hline
\tablecomments{\\p: power-law\\b: broken\\t: turnover\\ f: flat\\u: unclear}
 &   &  & \\
\end{tabular}
\label{tab_fits}
\end{table}

In this paper we have investigated whether the free-free absorption in the stellar wind can be the main reason why the spectra of PSR B1259$-$63 undergoes the observed evolution. We have modeled the stellar wind in the vicinity of the PSR B1259$-$63/LS 2883 binary system and created the computer simulation of PSR B1259$-$63 spectrum evolution. We have also presented implications of PSR B1259$-$63 case on classification of radio pulsar spectra in general. Additionally, in this paper we have made a detailed study of the PSR B1259$-$63 spectra using the flux measurements obtained during three orbital cycles \citep{johnston99, connors2002, johnston2005}. We have constructed radio spectra for different observing days and analysed their shapes taking into account the influence of interstellar scintillations. Let us note that previously the evolution was shown using the spectra averaged over the orbital phases. In addition to Paper I here we have studied the shapes of spectra for each observing day, in order to compare them to the modeled spectra that have been influenced by the free-free absorption in the hot stellar wind.

\section{PSR B1259$-$63 radio spectrum: detailed analysis}
As it has been mentioned above, a spectrum of the typical pulsar usually can be described in terms of a simple-power law in the form
\begin{equation}
S(\nu)=c\cdot\nu^{\xi},
\label{power_law}
\end{equation}
which is a good approximation except a frequency range below some critical cutoff frequency. We presume that a cutoff of a spectrum is related to the radio emission mechanism and is caused by a loss of coherence \citep{mgp00,glm04}.

However, there are also known more complicated spectra, e.g. turnover or broken-type. In the case of the broken-type spectrum it is necessary to fit the data to two power laws, which can be written as
\begin{equation}
S(\nu)=\left\{\begin{array}{ll} c_1\cdot\nu^{\xi_1}: & \nu\leqslant\nu_{b}\\ c_2\cdot\nu^{\xi_2}: & \nu>\nu_{b} \end{array}\right.
\label{broken}
\end{equation}
Here $\nu_{b}$ denotes the break frequency. The broken-type spectrum is usually defined as a spectrum fitted by two negative spectral indices and becoming steeper at higher frequencies. Spectrum with turnover, either at low or high frequency, can be modeled in a way proposed by
\citet{kuzmin2001}
\begin{equation}
S(\nu)=10^{ax^2+bx+c},\quad x\equiv\log_{10}{\nu},
\label{turnover}
\end{equation}
where $\nu_{p}=10^{-\frac{b}{2a}}$ denotes the peak frequency.

Function~(\ref{turnover}) was used in Paper I to analyse both the shapes of GPS and the evolution of PSR B1259$-$63 spectrum in various orbital phases. The authors, however, used the data with flux measurements at four frequencies for each given day except an orbital phase corresponding to 18 days prior to periastron passage. In this orbital phase, which is just before the eclipse, flux measurements were available only at three frequencies. Therefore, to improve results obtained in Paper I, we have created a database with all available flux density measurements for this pulsar. Using additional observational data, which were not taken into account previously, we have constructed the radio spectra of PSR B1259$-$63 for various orbital phase ranges and attempted to fit both functions (\ref{broken}) and (\ref{turnover}) to each spectrum. Then we have chosen those fits which most accurately reflect the  shapes of the spectra and used them to re-examine the spectrum evolution presented in Paper I. We have used the methods which are a combination of the methods used by \citet{maron2000} for radio spectra of about 280 pulsars and the methods used in Paper I for PSR B1259$-$63 spectrum evolution.

\begin{figure*}[!ht]
  \begin{tabular}{@{}cc}
   \includegraphics[height=5.6cm]{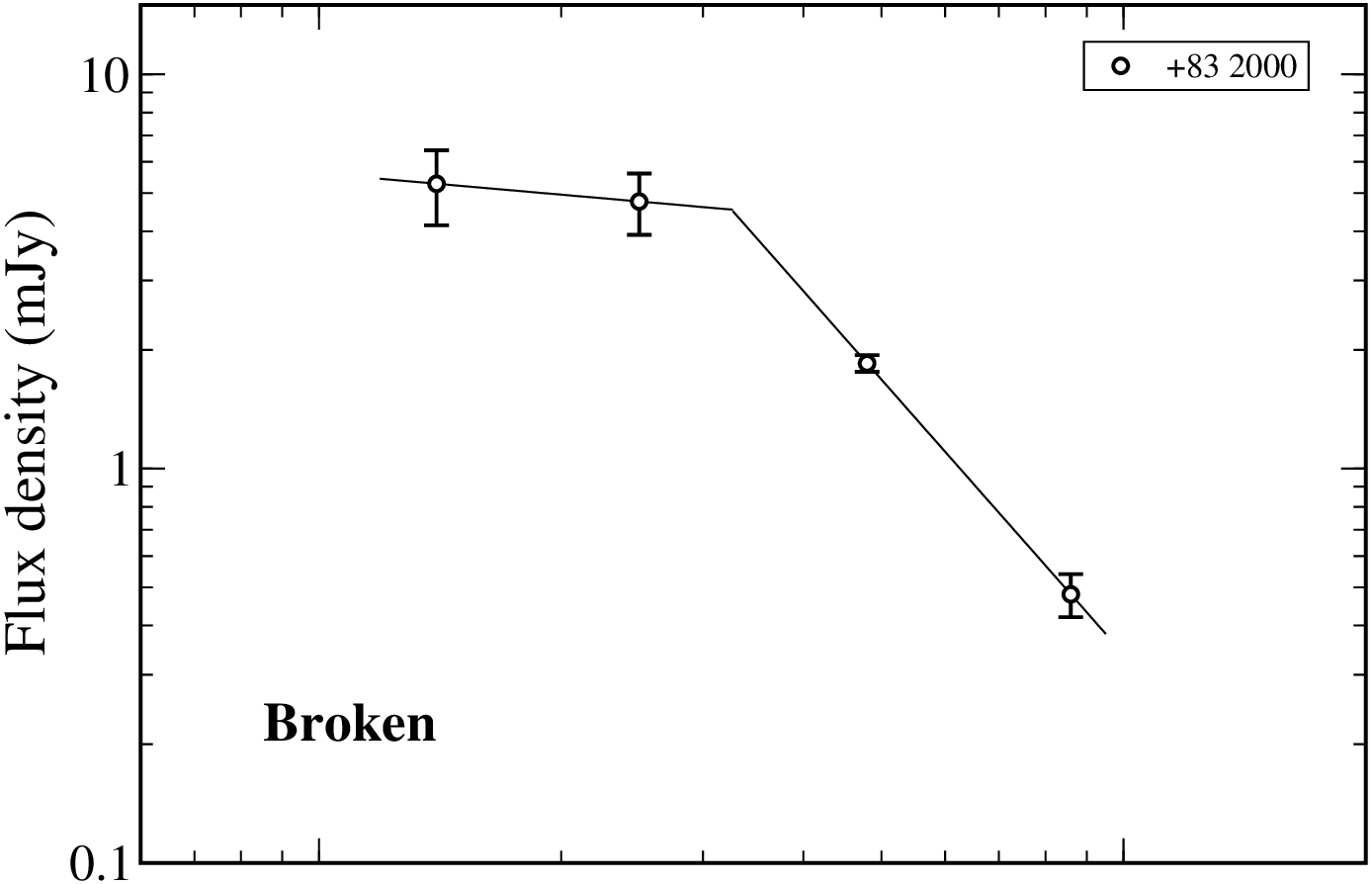}
   \includegraphics[height=5.6cm]{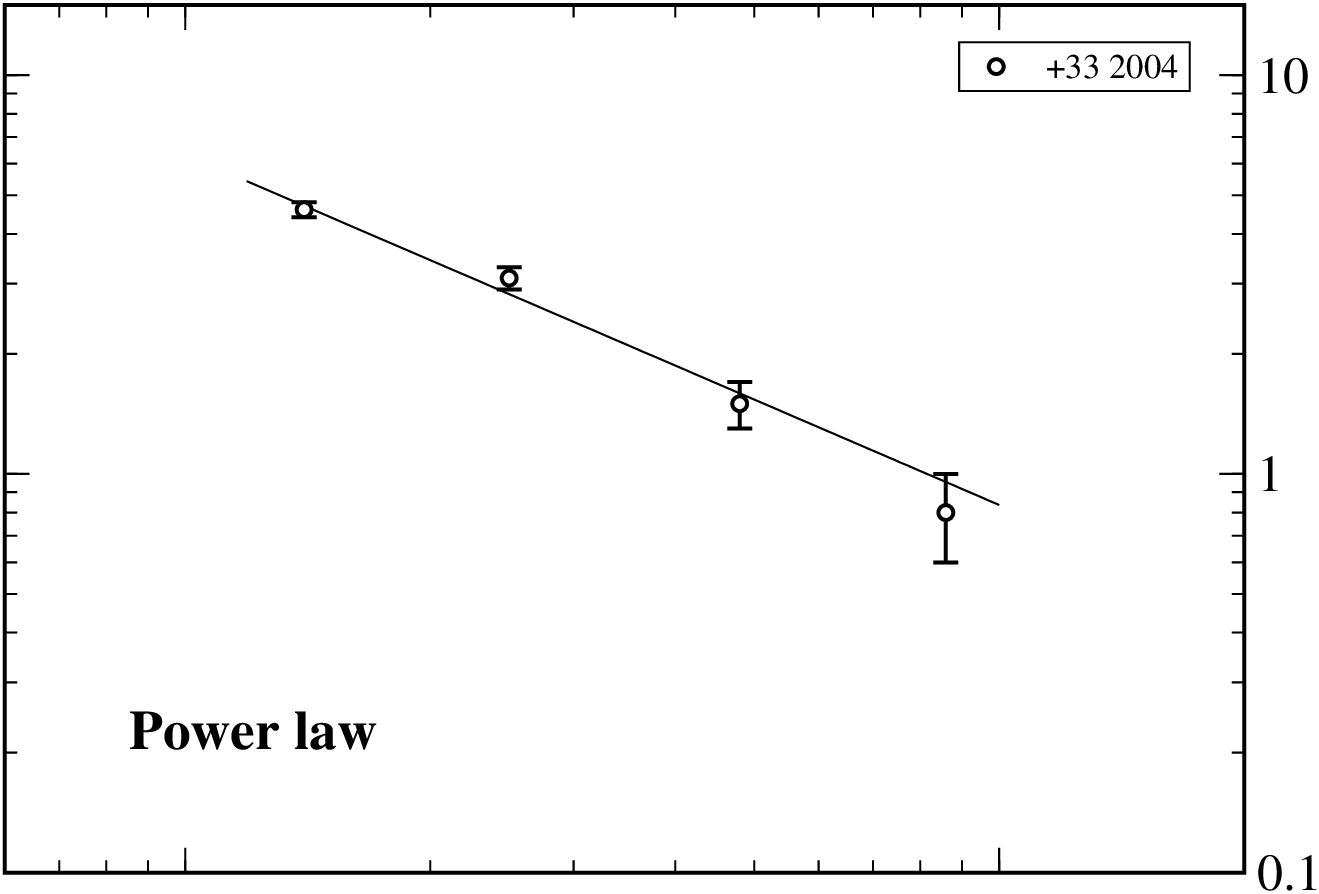}\\
   \includegraphics[height=6.2cm]{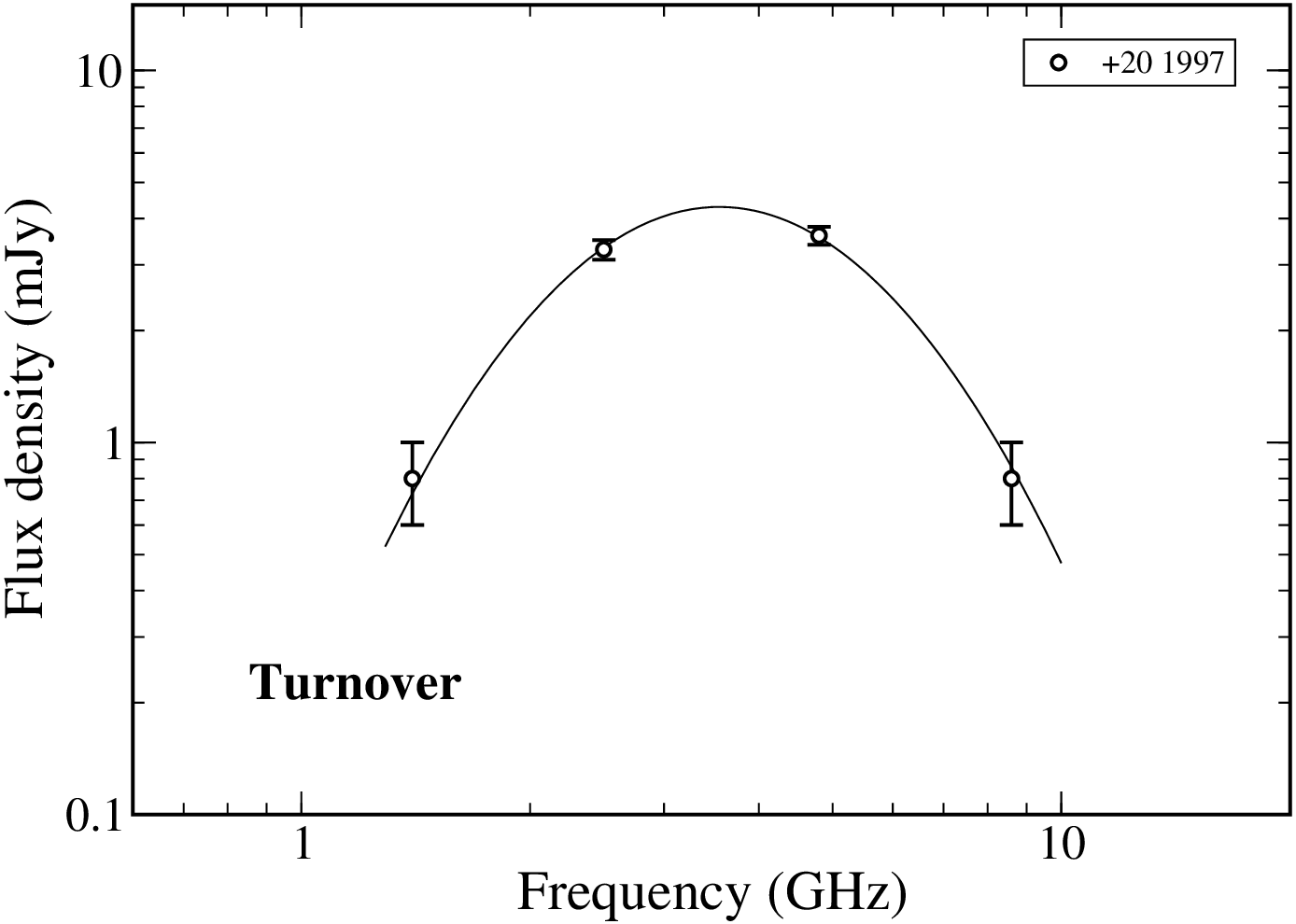}
   \includegraphics[height=6.2cm]{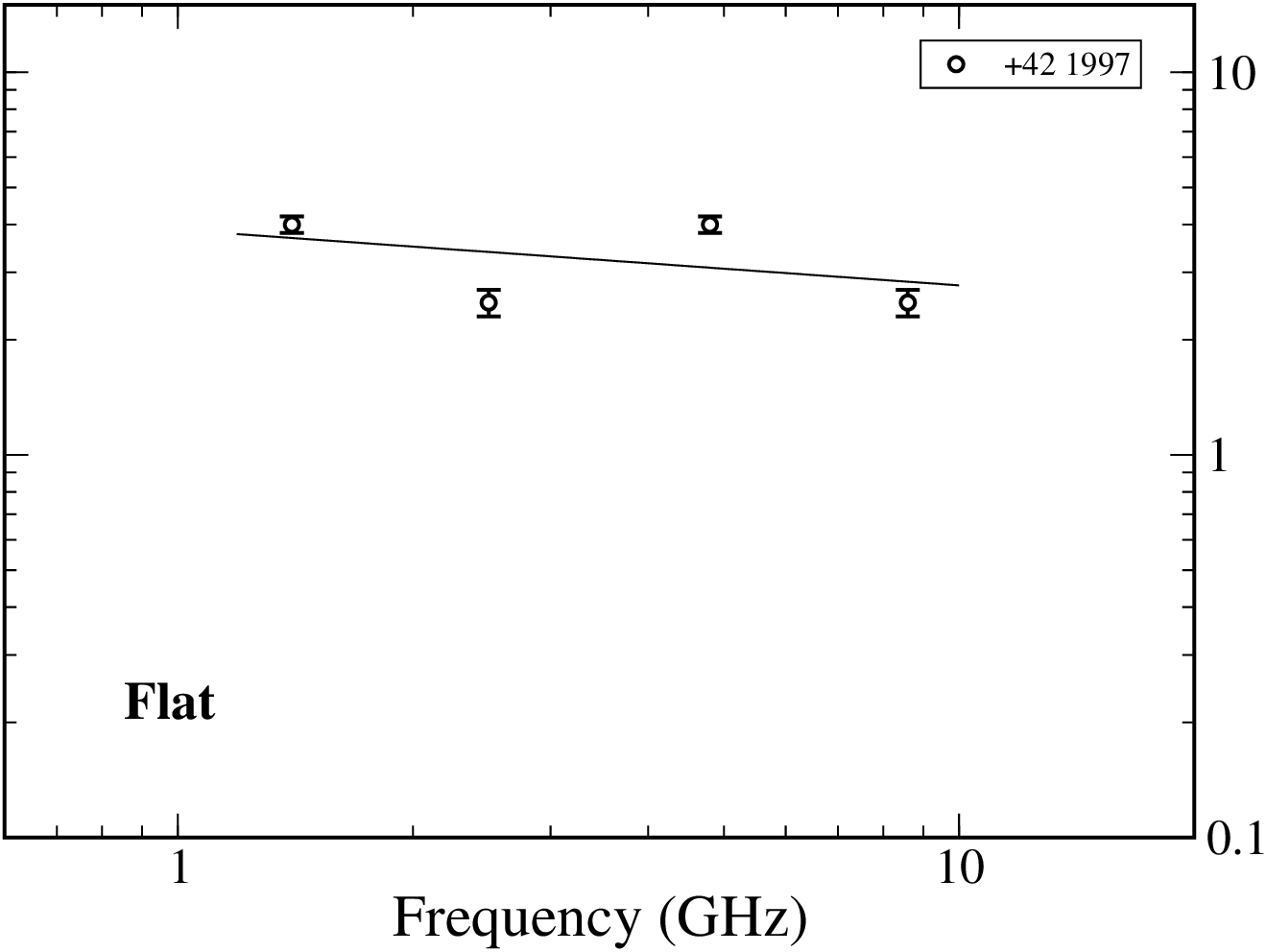}\\
   \end{tabular}
   \caption{The fits to the PSR B1259$-$63 spectra for chosen days (``-" and ``+" denote days before and after periastron passage respectively). Each panel shows a different type of spectra.}
  \label{detailed}
\end{figure*}

\begin{table}[!b]
  \caption{Best fits and parameters fitted to spectra presented on Fig.\ref{detailed}, for more details see Eq. (\ref{power_law}), (\ref{broken}) and (\ref{turnover}); ``+" denote days after periastron passage, year is an epoch of observation.}
  \begin{tabular}{@{}clcl@{}}
  \hline
   Day     &    Year     &  Best fit  & Parameters \\
\hline
 +33  & 2004 & power-law  & $d=6.3$, $\alpha=-0.88$\\
 +83  & 2000 & broken & $d_1=5.6$, $\alpha_1=-0.18$, \\
          &          &            &  $d_2=69.7$, $\alpha_2=-2.3$, \\
          &          &            & $\nu_{b}=$ 3.3 GHz\\
 +20  & 1997 & turnover & $a=-4.7$, $b=5.2$,\\
          &         &                & $c=-0.79$, $\nu_{p}=$ 3.6 GHz\\
 +42  & 1997 & flat & $c=3.9$, $\alpha=-0.14$\\
\hline
 &   &  & \\
\end{tabular}
\label{fit_days}
\end{table}

\subsection{Data analysis}
The fitting have been performed using an implementation of the nonlinear Marquard-Levenberg least-squares fitting algorithm. To choose the best fit we have analysed the results in a complex way.  At first, we have rejected the evidently inaccurate approximations and non-physical results, e.g. break at frequency below the lowest or above the highest frequencies. One cannot simply compare the values of $\chi^2$ to denote the best fit because there are available at most four data points. Otherwise, most of spectra would be classified as broken-type spectra. Let us note that based on an existing classification of spectra of pulsar radio emission we can distinguish three basic types of the spectra.

As mentioned above, we expect the broken-type spectra to have two negative spectral indices. This in turn leads us to classify the spectra showing positive spectral indices below some peak frequency as the turnover spectra. We also specify power-law spectra with a very small value (close to zero) of spectral index as the flat spectra. Some of the spectra show two different trends (for more details see Tab.~\ref{tab_fits}). Any of other existing cases are denoted as unclear. We have presented the best fits for each spectrum in Table \ref{tab_fits}.

\setcounter{figure}{1}

\subsection{Results}

In Paper I, there was presented a study of PSR B1259$-$63 spectra constructed for some orbital epochs. The authors also analysed published data on the pulsar radio fluxes measured during three periastron passages in: 1997, 2001 and 2004. While calculating the average flux densities in Paper I, however, only those data sets were used that contained the flux measurements at least for four frequencies in a given observing day. In the present paper we have extended the database by including all measurements published so far. Fig.~\ref{intervals} shows the spectra for various orbital phase ranges which can be used as a reference spectra when analysing the spectrum for a given observing day. It seems clear that the shape of the PSR B1259$-$63 spectrum depends on the orbital phase. Therefore, we can acknowledge that the spectrum of PSR B1259$-$63 undergoes evolution. In Paper I it was suggested that the free-free absorption in the stellar wind can be responsible for this effect. Moreover, our analysis confirms that the shape of spectra are not fully symmetric with respect to the periastron point (see Paper I). One can note that the flux density values on the left panel are generally smaller that those on the right panel of Fig.\ref{intervals}. This effect can be explained by influence of the stellar wind as the waves emitted before the periastron passage travel longer distance through the wind (see Fig.~4 in Paper I). Thus, they are affected by the free-free absorption stronger than the waves emitted after the eclipse. We can also confirm that the peak (or in some cases break) frequency depends on the orbital phase.

Comparing the results of fitting presented in Tab.~\ref{tab_fits} one can notice that the shape of spectrum obtained in a single day of observation is generally consistent with the spectrum obtained by averaging over the corresponding orbital phase interval. One can also note the GPS feature more or less dominates during the orbital phases before the eclipse. It is worth mentioning that the first sign of a typical spectrum (i.e. broken or power-law) appears as early as 20 days after the periastron passage. The flux density at the given frequency apparently changes with orbital phases. When the pulsar is close to periastron, the flux generally decreases at all observed frequencies and the most drastic decrease is observed at the lowest frequency. Moreover, one can distinguish all types of radio pulsar spectra, including a flat spectrum (see Fig.~\ref{detailed}). Close to the periastron point the spectra of PSR B1259$-$63 resemble those of the GPS pulsars, while for the orbital epochs further from the periastron point the spectra are more consistent with typical pulsar spectra (i.e. power-law or broken).

\section{PSR B1259$-$63 radio spectrum evolution: simulation}
\label{model}
The variation of PSR B1259$-$63 spectrum shape is believed to be caused by two physical processes: the free-free absorption and the cyclotron resonance (Paper I). In our computations only the first mechanism is considered as we have neglected the presence of the stellar equatorial disk. In order to examine whether the free-free absorption in the hot stellar wind is able to account for the observed spectral evolution of PSR B1259$-$63 we used a simplified model for the hot stellar wind. It should be noted that we do not pretend to explain in details the behaviour of the pulsar spectra using this model.

\subsection{The model}
In the model we assume that the stellar wind has a spherical symmetry, the electron density
decreases as $1/r^2$, the wind speed and the electron temperature are constant. We use the density of the stellar wind at 1~AU from the central object as the reference value. In such a model one can, of course, expect results to be symmetric with respect to the periastron passage. Optical depth depends on thickness of an absorbing medium as well as on its properties, and therefore it is changing during the pulsar orbital motion. However, as the pulsar gets closer to periastron the absorption increases since there also increases the fraction of the pulsar's line-of-sight that lies inside the stellar wind. Another factor that increases the absorption is the electron density which also increases when the pulsar's line-of-sight approaches the Be star.

One more important factor affecting a resulting spectrum of a pulsar is the shape of its intrinsic spectrum. In our simulations we assume that the intrinsic spectrum can be described by a single power-law function. Since there is no way we can observe the intrinsic spectrum, we have used two different spectral indices in our simulations, namely $\xi = -0.8$ and $\xi= -1.3$. The first one, $\xi = -0.8$, is the spectral index obtained from averaging of all the available data, and the second one, $\xi= -1.3$, is the average spectral index based on measurements performed only on 5~GHz and 8~GHz frequencies. We believe that the latter ($\xi= -1.3$) is more realistic, since higher observing frequencies are supposed to be less affected by absorption, almost regardless of the pulsar's orbital phase.

We have calculated the pulsar flux absorption using the radiative transfer equation, where the optical depth is given by the following formula \citep{sieber73}:
\begin{multline}
\tau_{\nu}= 3.014\cdot 10^{-2} \left( \frac{T_{e}}{\mathrm{K}}\right)^{-\frac{3}{2}}
\left(\frac{\nu}{\mathrm{GHz}}\right)^{-2}
\left(\frac{EM}{\mathrm{pc \cdot cm^{-6}}}\right) \\  \left[ \mathrm{ln} \left( 4.955 \cdot 10^{-2} \frac{\nu} {\mathrm{GHz}}\right)^{-1}+1.5 \mathrm{ln} \left(
\frac{T_{e}}{\mathrm{K}}\right)\right].
\end{multline}
Here $T_{e}$ denotes the electron temperature and an emission measure $EM$ is defined as
\citep{rohlfs2004}
\begin{equation}
\frac{EM}{\mathrm{pc}\cdot \mathrm
{cm^{-6}}}=\int_{0}^{\frac{s}{\mathrm{pc}}}\left(\frac{N_{e}}{\mathrm{cm^{-3}}}\right)^{2}\mathrm{d}\left(\frac{s}{\mathrm{pc}}\right),
\end{equation}
and $N_{e}$ denotes the electron density.

The simulated pulsar orbit has a semi-major axis of $a=4.41$ AU and the eccentricity of  $e=0.87$. The inclination of the orbit is set to $i=36^{\circ}$ and the orbital period is assumed to be $P=1237$ days. For our initial approach we have positioned the orbit in such a way, that the projection of the apse line is pointing towards the observer and the periastron is the furthest away point of the orbit ($\omega = 90^{\circ}$).

\subsection{Results}
\begin{figure*}[!ht]
  \begin{tabular}{@{}cc}
   \includegraphics[height=6.3cm]{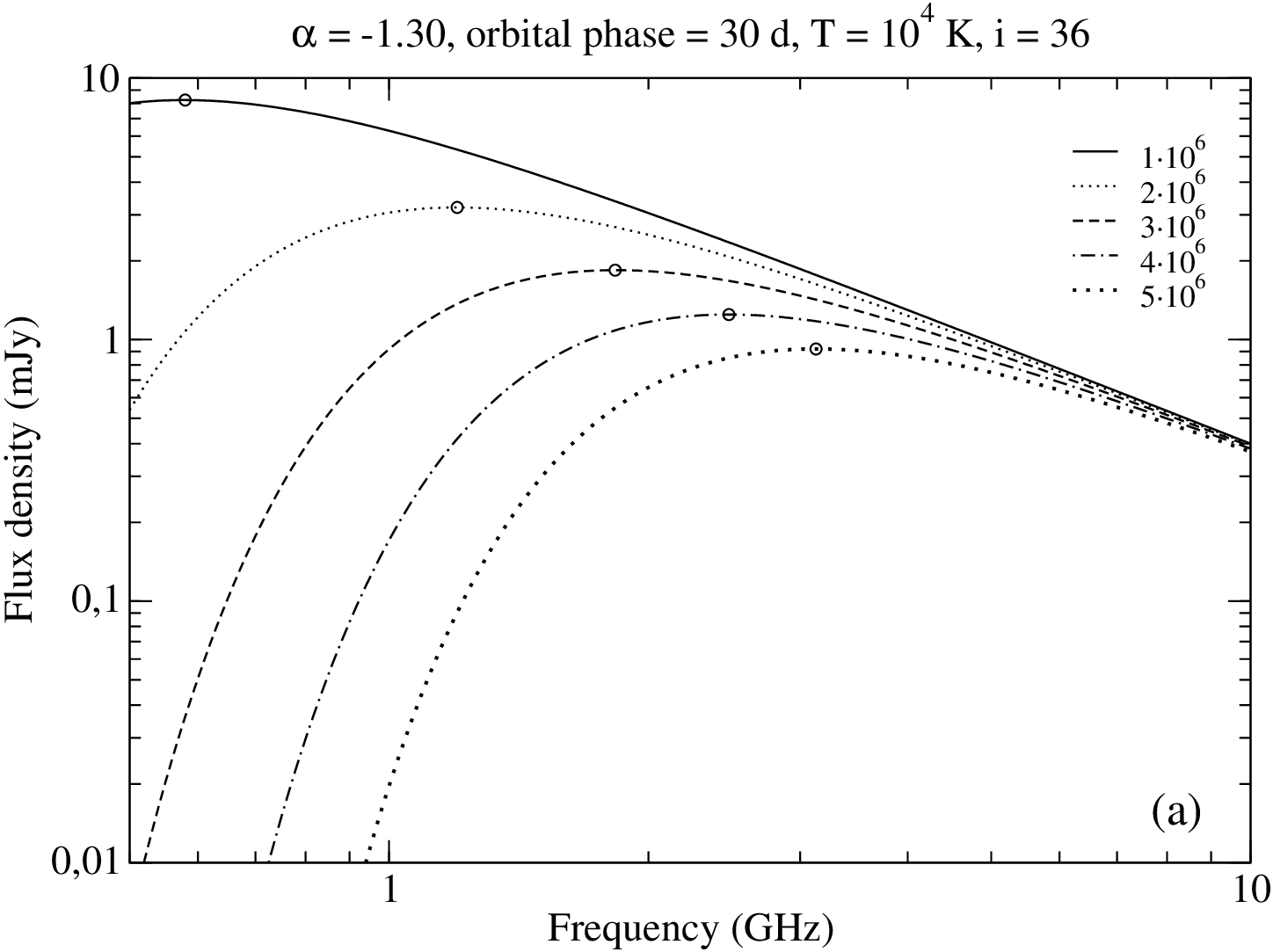}\hspace{0.2cm}
   \includegraphics[height=6.3cm]{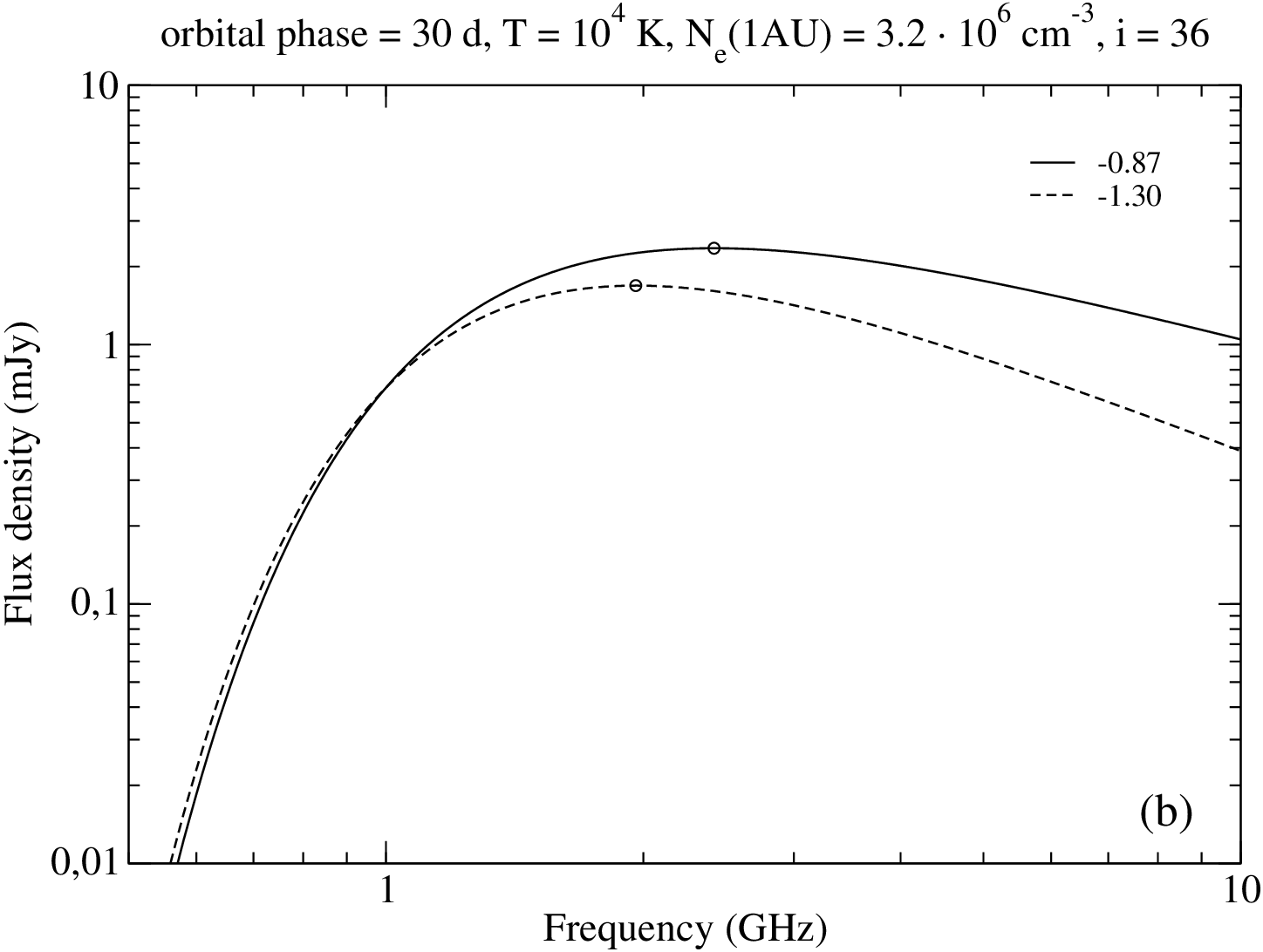}\\
      \vspace{0.1cm}\\
   \hspace{0.1cm}\includegraphics[height=6.3cm]{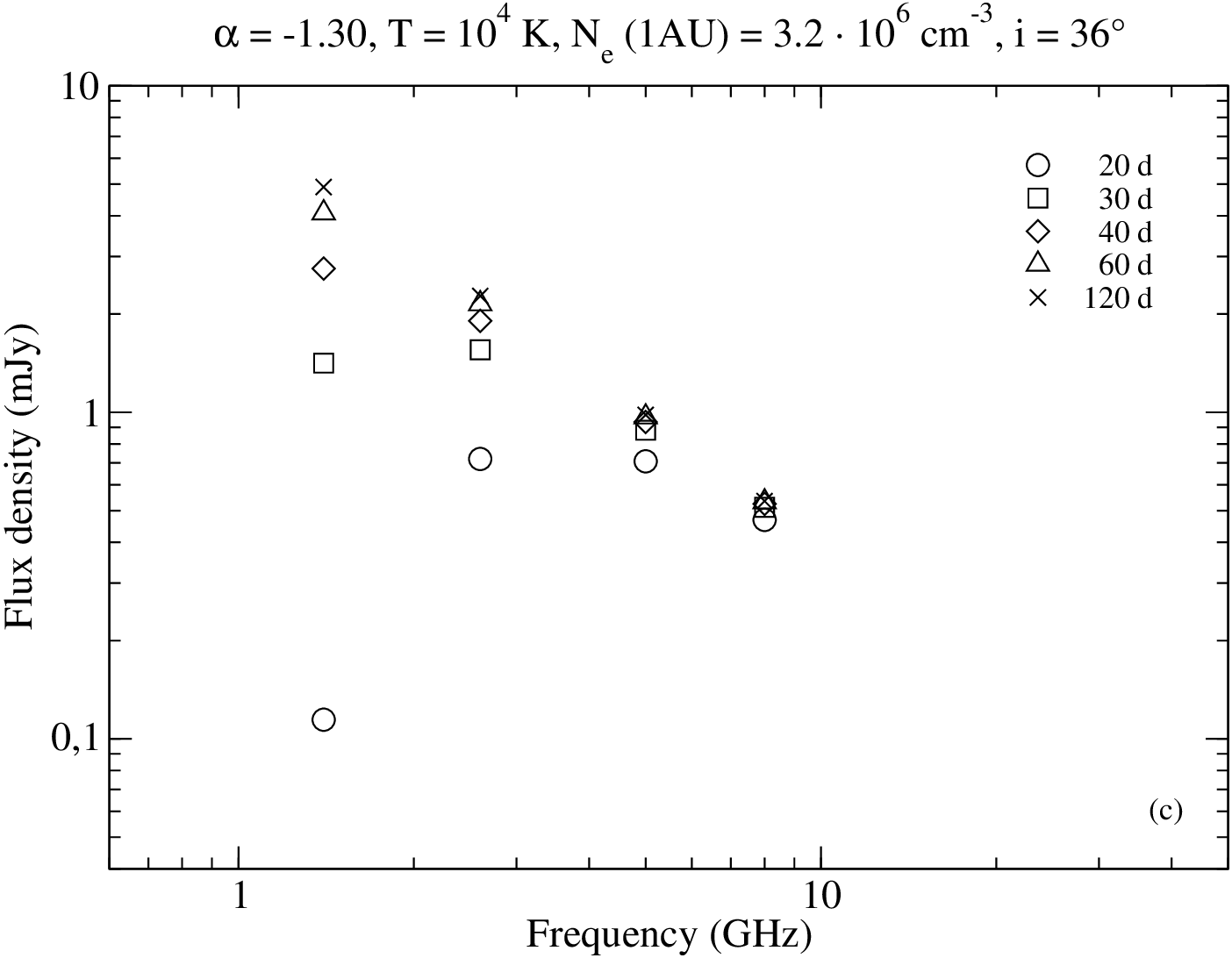}\hspace{0.4cm}
   \includegraphics[height=6.3cm]{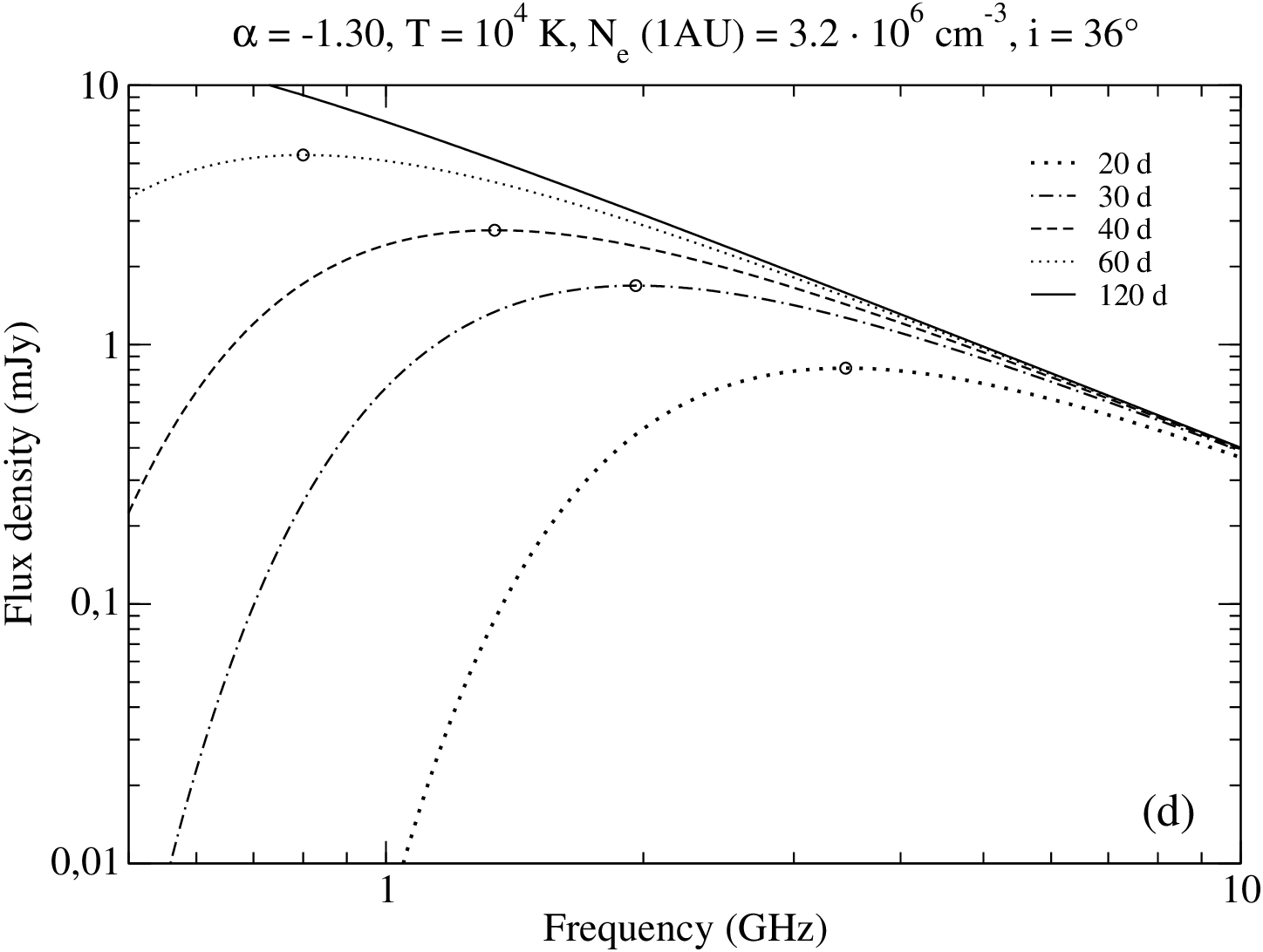}\\
   \end{tabular}
   \caption{Spectra resulted from modelling the influence of the free-free absorption for the PSR B1259$-$63 radio emission. The orbit inclination $i$ is equal 36$^\mathrm{o}$ and the electron temperature $T_{e}$ is equal $10^{4}\mathrm{K}$. The open dots denote the peak frequencies. 
   (a) Pulsar spectra for the different values of the star wind electron density. The values of the spectral index, the star wind electron temperature and the day after periastron passage are fixed. The optical depth increases with electron density. The same results would be obtained for a decreasing electron temperature at a fixed electron density value.
   (b) Flux density spectra for two different values of the spectral indices at the fixed values of the star wind electron temperature and density. (c) Flux density values for chosen frequencies (1.4, 2.6, 4.8 and 8.4 GHz) for the chosen days after periastron passage and for fixed values of the spectral index, the star wind electron temperature and density. (d) Pulsar spectra for the chosen days after the periastron passage and for the fixed values of the spectral index, the star wind electron temperature and density}
  \label{evolution}
\end{figure*}
To estimate the electron density that is needed to explain the observed spectral shapes we have assumed that the pulsar is located at a fixed position (corresponding to 30 days after the periastron passage) and modeled the absorption using various values of the electron density (~$N_e$~) at the chosen reference distance, i.e. at 1~AU. The results of our simulations are presented in panel (a) of Fig.~\ref{evolution}, where the different line types correspond to different values of $N_e$. As expected the amount of absorption at all observed frequencies increases with increasing value of $N_e$. Along with the increase of $N_e$ the peak frequency (the open dots in Fig.~\ref{evolution}) shifts towards the higher observed frequencies as the absorption is more significant at lower frequencies, consequently the total observed flux steadily decreases. As one can see in panel (a) of Fig.~\ref{evolution} that the stellar wind with the electron density of the order of a few million particles per cm$^3$ (at the distance of 1~AU from the star) is enough to explain the observed behaviour of the spectra. Panel (b) of Fig.~\ref{evolution} shows the apparent change of the out-coming spectrum for two intrinsic spectral indices, $\xi = -0.8$ and $\xi= -1.3$. As one can see, using the steeper intrinsic spectrum yields the sharper shape of the observed spectrum and slightly shifts the peak frequency towards the lower observing frequencies.

As the next step of our simulation we have used the intrinsic spectrum with the spectral index of $\xi= -1.3$ and have calculated the shape of the spectrum for various orbital phases. Like the real observations we denote them by number of days pre or post the periastron passage. Panels (c) and (d) of Fig.~\ref{evolution} demonstrate our results. Panel (d) shows the spectra obtained during the orbital phase starting from 20 days after the periastron passage\footnote{The same effect can be obtained prior to periastron as our model is symmetric and does not include the real value of $\omega$.} up to 120 days after, when the outgoing spectrum is almost the same as the intrinsic single power-law pulsar spectrum (since there is almost no absorption at all). As one can clearly see the peak frequency in the spectrum shifts significantly towards lower frequencies as the pulsar moves away from its companion. Panel (c) of Fig.~\ref{evolution} shows the simulated radio fluxes at the four observing frequencies, i.e. the frequencies which are used to perform the real observations.

As one can see the simulated flux densities corresponding to the orbital phase 20 days after the periastron passage (open circles) can be clearly identified as a GPS type, although we admit that it is not as symmetric as the real spectrum from that phase (see the bottom left plot of Fig.~\ref{detailed}). The spectra from day 30 after the periastron passage (open squares) would be identified either as a GPS or a broken-type, and from day 40 -- possibly marginally a broken type or a single power-law spectrum. The spectra from later orbital phases (60 and 120 days, showed by triangles and crosses, respectively) is hard to distinguish from a regular pulsar spectrum.

We believe that this simple model shows its ability to simulate some basic properties of the observed spectra, although the model clearly requires further investigation. Obviously, in its current stage the model is not able to recreate the observed asymmetry between the spectra collected prior to and after the periastron passage. To model the asymmetry we need to introduce the real geometry of the system (namely the real value of the ascending node angle $\omega$). Also, starting from about 40 days after the periastron passage our simulated spectra (within the chosen frequency range) tend to show almost no absorption, which is not true, as the observed spectra show some absorption (see Fig. ~\ref{intervals}). This should  be caused by the real geometry of the stellar wind from a Be star which is bi-modal, constituting from the slow equatorial and fast polar winds. Hence the wind distribution is far from isotropic. Let us note, that in our simulation we assumed the wind being spherically symmetric. Finally, our simple model does not include possible cyclotron resonance effects that may happen when the pulsar comes near the stellar disk.

\section{Discussion}

\subsection{Influence of interstellar scintillations on PSR B1259$-$63 spectrum}

Pulsars are generally known to be stable radio sources although the pulsar signal is affected in a number of ways by the interstellar medium \citep{lk}. Thus, the measured flux density varies due to diffractive and refractive scintillation effects. Scintillation studies are extremely successful in determining some properties of the interstellar medium and positioning of the dominant scattering screen. PSR~B1259$-$63 is a pulsar with a relatively high dispersion measure, therefore, the scintillation is in the strong regime. Thus we definitely have to take into consideration both refractive (RISS) and diffractive (DISS) scintillations while analysing the spectra.

To estimate the characteristic timescales $\Delta t_{\mathrm{RISS}}$ and $\Delta t_{\mathrm{DISS}}$ of scintillation we have used observations done by \citet{scin98}. They observed PSR B1259$-$63 at 4.8 GHz and 8.4 GHz during 46 sessions between August 1993 and February 1997, however, excluded from their analysis the data taken during the periastron passage. The observations were typically between 30 and 100 minutes long.

Based on their results we have estimated the values of $\Delta t_{\mathrm{DISS}}$ to be ranging from 40s at 1.4~GHz to 360s at 8.4~GHz which suggests that the diffractive scintillations should not affect the average flux measurements as the observing sessions we use are usually 4 hours long. In contrary the refractive scintillations may play a significant role, as the estimated refractive timescales $\Delta t_{\mathrm{RISS}}$ vary from 12~hours at~8.4~GHz to more than 20~days at~1.4~GHz. At lower frequencies, however, the modulation index is relatively small which means that the uncertainty of flux measurement is also relatively small. On the other hand, observations at higher frequencies may be affected strongly by the refractive scintillations, as the  modulation index is higher and the typical duration of an observing session (4 hours) is still shorter than $\Delta t_{\mathrm{RISS}}> 12$ hours. We can definitely point out at least a couple of high frequency observations that are affected by refractive scintillations. Definitely, this is the case for the orbital phase range from +113 to +186 days (see Tab.~\ref{tab_fits}), which includes the data obtained during only three sessions. Thus, the spectrum obtained for this orbital phase range (see the dashed line in the right panel of Fig.~\ref{intervals}) may differ essentially from the real and/or simulated (see the spectrum represented by solid line in the bottom right panel (d) of Fig.~\ref{evolution}) spectra.

It should be mentioned, that despite the above example, it is always profitable to average data obtained during few neighbouring orbital phases to get more reliable estimate of the spectrum. However, let us note, that the clear signature of the spectral evolution is recognisable even by classification of spectra obtained during each session (see column 3 in Table \ref{tab_fits}), as the spectra close to periastron seem to be classified mostly as ``turnover", while the spectra far from periastron preferably are classified as ``broken" or ``power-law".

\subsection{Conclusions from the simulation of the spectrum evolution}

As it was stated in Paper I, the case of PSR B1259$-$63 can play a crucial role in understanding the main factors that affects the observed spectra. We believe that it is important to investigate how and why the changing environment causes the spectrum evolution. As an initial approach to study this problem we have constructed a simplified model of the system which includes the thermal absorption in a spherically symmetrical stellar wind (see Section~\ref{model}). Using this model we are able to recreate the basic properties of the PSR B1259$-$63 spectrum evolution, although the model is far from being perfect and does not explain some peculiarities of the spectrum variations. However, we have estimated that the electron density has to be of the order of a few million particles per cubic cm (at the reference distance of 1~AU) in order to get a necessary absorption. 
Since the direct measurements of the electron density around Be stars are not available (this quantity is almost impossible to measure) we decided to roughly estimate the expected value based on a simple model of the stellar wind. If we assume that the entire loss of the mass by a star is due to the wind, then a value of $\dot{M}$ (usually expressed in solar masses per year) can be easily translated to the number of escaping particles per year (or per second), if one assumes that all the mass escapes the star in form of protons. Next, we assumed that the number of escaping electrons is roughly the same as the number of protons (since electrons are much lighter they almost do not contribute to the stellar mass loss). Knowing the number of electrons escaping the star every second we can calculate how many of them will pass through an unit area (i.e. the particle flux) at any distance from the star, since we simply need to divide the number of escaping electrons by the area of the corresponding sphere.  Next, if we assume some velocity of the stellar wind (we used 1000 km/s in our calculations), one can find the number density of the particles corresponding to the calculated amount of the particle flux.

Using the above reasoning we estimated, that for the density of the escaping electrons to reach a value of a few millions of particles per cm$^3$ one needs to adopt the stellar mass loss of the order of $10^{-9} M_{\odot}/{\rm year}$. This mass-loss rate seems to be reasonable when comparing to values reported for this object by \citet{johnston96} and \citet{shannon2014} which is $\dot{M} \sim10^{-8} M_{\odot}/{\rm year}$. The observed value is by an order of magnitude larger than our estimate based on the absorption, but one has to remember that most of the stellar mass escapes a Be star though a slow equatorial wind, and in our model of absorption the pulsar spectra will be affected only by the weaker polar wind.

It is obvious that the model requires further modifications to simulate all the characteristics of the observed spectrum evolution. Nevertheless we believe that this model, even in its current simplified state shows that the thermal free-free absorption occurring in the stellar wind is a viable explanation for the pulsar spectra shapes and their evolution.

The next step should be incorporation of the second wind component into the model. The equatorial wind could influence especially the spectrum observed during the orbital phases before and after the eclipse (approximately $\pm$~20 days). At the moment it is difficult to estimate the parameters, i.e. density, temperature and velocity, of this component.

It is also important to estimate the electron density by independent (from absorption) method, using estimations of the dispersion measure variability that corresponds to various orbital phases \citep{johnston2005}. In our model we have estimated the optical thickness of the environment that guarantees the observed absorption. The depth, however, depends on both the temperature and the density of electrons, and the independent estimation of the electron density allows to separate influences of these parameters. Preliminary estimations show that the densities that we obtain from our model are consistent with the observed DM variations\footnote{For specific values of dipersion measure see Table 2 in \citet{johnston96}, Figures 1 and 2 in \citet{johnston2001}, Table 3 in \citet{connors2002}, Figure 2 in \citet{wang2004} and Table 1 in \citet{johnston2005}. }. In order to calculate accurately variations of DM caused by the hot stellar wind we need to use the dispersion law in the hot plasma.

It is clear that more radio observations of PSR B1259$-$63 corresponding to various orbital phases are necessary. Particularly, it is very important to measure the flux density near the apastron passage to obtain the intrinsic spectrum of the pulsar. Moreover, additional observations at as many as possible orbital phases will help us to take into account the influence of interstellar scintillations, as its impact on the shape of observed spectra can be critical (see section 4.1).

\section{Summary}

The same time the detailed analysis of the PSR B1259$-$63 spectra for observing days reveals the appearance of all types of spectra. Furthermore, the fits presented in Tab.~\ref{tab_fits} show significant differences between the spectra obtained before and after eclipse. The results lead us to the conclusion that the PSR B1259$-$63 spectrum varies with the changes of the pulsar environment. Our simulations based on the simplified model of the stellar wind have shown that the free--free absorption can be a cause of an observed variety of spectral types. Assuming reasonable parameters (an electron density and temperature, the wind speed, etc.) of the stellar wind the simulations have shown that the obtained spectrum evolution is consistent with the observational data. The results also support the hypothesis that external factors can have a significant impact
on the observed radio emission of a pulsar.

We believe that the case of PSR B1259$-$63 can be treated as a key factor to our understanding of not only the GPS phenomenon (observed for the solitary pulsars with interesting environments) but also other types of untypical spectra as well (e.g. flat or broken spectra). This in turn would suggest, that the appearance of various non-standard spectra shapes in the general population of pulsars can be caused by peculiar environmental conditions.

\section*{Acknowledgments}
This paper was supported by the grants DEC-2012/05/B/ST9/03924 and DEC-2013/09/B/ST9/02177 of the Polish National Science Centre. MD was a scholar within Sub-measure 8.2.2 Regional Innovation Strategies, Measure 8.2 Transfer of knowledge, Priority VIII Regional human resources for the economy Human Capital Operational Programme co-financed by European Social Fund and state budget.

\end{document}